**The use of cost information when defining critical values for prediction of rare events using logistic regression and similar methods.**

Berk *et al.* (2009*)* present an approach they have themselves developed: "statistical learning by random forests" (SLRF) for forecasting rare events that are difficult to predict. However, they unfairly stigmatise logistic regression when used for this purpose. Their objection is that very few subjects are identified by logistic regression as more likely than not to be cases (posterior probability > 50%); and they imagine that nothing can be done to improve this.

In their chosen example, (forecasting murder within a population of probationers and parolees) they balance a rare but very serious outcome for a false negative (murder) against the more common, but relatively less serious consequence of a false positive (holding someone either charged or convicted of a crime in prison without need). Balancing a rare and serious possibility against a more common and less serious one is a familiar problem in many other situations, such as the prediction of rare diseases.

They approach the problem by taking the relative costs of forecasting errors into account. As I demonstrate, this can be done for any prediction method that gives an estimated probability of a future event. All that is needed is to work out the probability at which the likely cost (defined as *cost x probability*) of a possible false negative is exactly equal to that of a possible false positive. This gives the relevant cutpoint and all subjects with probability of disease greater than this have a positive test result. All standard methods of logistic regression will give the log-odds and hence the predicted probability of a positive outcome for every subject:

> *relative cost = (cost of false negative)/(cost of false positive)*
> *probability cutpoint = 1/(1+ relative cost)*
> *predicted probability = exp(log odds)/(1+exp(log odds))*

Although this approach is straightforward, I am unable to point to any examples where it has been used in practice. The idea of combining costs and probabilities will be familiar to Bayesians, but its case-by-case use for decision-making is unusual.

In the training sample of 30,000 subjects the event occurs only 322 times (1.1%). With this prevalence, any predictive method would struggle to identify more than a handful of cases with a probability of 50% or more. Indeed, the likelihood ratio of a positive test (LR+) would have to be at least 46.6 to convert 1.1% prevalence to a posterior probability of 50%. Their preferred model (SLRF) achieves a LR+ of only 7.16, giving a post-test probability of 12% (137/1,764). By their own rules, every subject should be treated as negative.

In this case, the cost ratio is set by Philadelphia's Adult Probation and Parole Department at 10:1. This should of course be thought of as cost to society, rather than simply financial cost. A homicide or attempted homicide committed while on parole or probation is regarded as ten times as serious as holding someone in prison without need. The authors assume that this ratio is properly used because they have approximately 10 times as many false positives as false negatives; but this is not the case. All errors are bad. The best possible decision needs to be made separately for each individual.

Using the formulae above, the cutpoint is at 1/(10+1) or 9.09%, and any subject with a posterior probability of recidivism (specifically homicide or attempted homicide) equal or greater than this would be treated as a high-risk. A 1/11 chance of a cost of 10 for a false negative is balanced by a 10/11 chance of a cost of 1 for a false positive. The likely cost is 10/11 of the cost of a false positive in each case.

It is possible to measure the performance of any model in units of the cost of one false positive. By this criterion, SLRF performs poorly in both the training and validation sets. In the training set the total cost is 1,764 + 10x185 = 3,614, outperformed by the poorly calibrated logistic regression model they use as a benchmark: total cost 321x10 +1 = 3,211. In the validation set (prevalence 348/33190), SLRF fails to identify 198 future homicides, and misclassifies 2,193 non-recidivists; total cost 2193+10x198 = 4,173 times the cost of a false positive. The simpler "method" of treating everyone as negative, and letting them all go free, would give a substantially lower total cost of 3,480 times the cost of a false positive. However, I am not actually suggesting that as public policy.

In this example, the cost ratio may be set too low. With a cost ratio of 15:1, the SLRF model would be as good as treating everyone as low risk; with one of 100:1, it would reduce costs by 38%. This sort of saving in costs to society would justify the positive reaction that has resulted from using the model.

It would be interesting to compare the confusion diagrams for the logistic model, based on a cutpoint at 9.09% with that from Berk *et al.*'s preferred method. However, this would only be a fair comparison if every attempt was made to fit the best possible logistic regression model; as this was done with SLRF.

In particular, I would be interested to see the use of a stepwise method of variable selection, based perhaps on maximising the Akaike Information Criterion (AIC). Continuous and ordered variables (such as current age, age at first offence, and number of offences) could be fitted using "traffic light" dummies, defined (for instance) as (current age>=16), (current age>=17), (current age>=18) etc. With this system, the exclusion of a single dummy will have the effect of combining adjacent groups. In addition, first order interaction terms might be used, as there are a relatively large number of cases. AIC is implemented within Stata by the -swaic- command (Wang 2000, Wang & Steichen 2005). However, if AIC is not available, stepwise regression with a relatively modest inclusion criterion (p<0.1) might be used.

Other approaches to the logistic regression are certainly possible; these are simply the ones that I myself would prefer.

There are several advantages to a logistic regression approach: the software is generally available. The modelling is relatively fast and simple once the dummy variables have been defined. The model terms may reveal something about the causes of the bad outcome (whether recidivism or disease). The model can be implemented as a simple formula and made generally available in a standard spreadsheet. If there is a change in the relative importance of false positives and false negatives, it is easy to allow for this.

Paul Seed
King's College London, Division of Reproduction and Endocrinology
St Thomas' Hospital, Westminster Bridge Road, London SE1 7EH, UK
E-mail: paul.seed@kcl.ac.uk